\documentclass[11pt,a4paper]{article}
\pdfoutput=1

\usepackage{amsmath}
\usepackage[T1]{fontenc}

\usepackage{jheppub}
\usepackage{psfrag}
\usepackage{slashed}
\usepackage{cancel}
\usepackage{lscape}

\usepackage{caption}
\usepackage{array}
\usepackage{graphicx}
\usepackage{subcaption}
\usepackage{multirow}

\usepackage[utf8]{inputenc}

\usepackage[utf8]{inputenc}
\usepackage{graphicx}
\usepackage{slashed}
\usepackage{color}
\usepackage{amsmath}
\usepackage{amssymb}

\definecolor{mygreen}{rgb}{0,0.7,0}

\usepackage{multirow}
\usepackage{tabularx}
\newcolumntype{C}[1]{>{\hsize=#1\hsize\centering\arraybackslash}X}%

\newcolumntype{Z}{r<{\hspace{3mm}}}

\def\nn{\nonumber \\ }

\def\la{\langle}
\def\ra{\rangle}
\def\spA#1#2{\la#1#2\ra}
\def\spB#1#2{[#1#2]}

\DeclareMathOperator{\tr}{\rm tr}

\def\trp{\tr_+}

\def\eps{\epsilon}

\def\fbox#1{F^{(#1)}_{\mathrm{box}}}
\def\lh{\hat{L}}
\def\li#1{\mathrm{Li}_{#1}}

\newcommand\cibp{c^{[{\rm IBP}]}}
\newcommand\camp{c^{[{\rm A}]}}
\newcommand\cir{c^{[\rm IR]}}
\newcommand\cfin{c^{[{\rm F}]}}
\newcommand\cpfunc{c^{[\rm f]}}
\newcommand\cintegrand{c^{[\partial{\rm A}]}}

\preprint{{IPPP/18/101, ZU-TH 44/18}}

\title{Analytic helicity amplitudes for two-loop five-gluon scattering: the single-minus case}

\author[a]{Simon Badger,}
\author[b]{Christian Br\o{}nnum-Hansen,}
\author[a]{Heribertus Bayu Hartanto,}
\author[c]{Tiziano Peraro}
\affiliation[a]{
Institute for Particle Physics Phenomenology, Department of Physics, Durham University, Durham DH1
3LE, United Kingdom%
}
\affiliation[b]{
Higgs Centre for Theoretical Physics, School of Physics and Astronomy, The University of Edinburgh, Edinburgh EH9 3JZ, Scotland, UK%
}

\affiliation[c]{
Physik-Institut, Universit\"at Z\"urich, Wintherturerstrasse 190, CH-8057 Z\"urich, Switzerland
}

  \emailAdd{simon.d.badger@durham.ac.uk, bronnum.hansen@ed.ac.uk, heribertus.b.hartanto@durham.ac.uk}

\abstract{We present a compact analytic expression for the leading colour two-loop five-gluon
amplitude in Yang-Mills theory with a single negative helicity and four positive helicities. The
analytic result is reconstructed from numerical evaluations over finite fields.  The numerical
method combines integrand reduction, integration-by-parts identities and Laurent expansion into a
basis of pentagon functions to compute the coefficients directly from six-dimensional generalised
unitarity cuts.}


\begin{document}
\maketitle
\flushbottom

\section{Introduction \label{sec:introduction}}

The increasing need for high precision predictions for Standard Model processes at Hadron colliders
has set out a priority list of new perturbative calculations required to keep theoretical
uncertainties in line with experimental errors~\cite{Bendavid:2018nar}. A large category of these
processes requires unknown two-loop $2\to3$ scattering amplitudes. Owing to the high degree of
difficulty of these calculations, recent years have seen an increasing effort to find new techniques
capable of providing these results.

Due to the success of automated algorithms for the numerical computation of one-loop amplitudes,
there has been considerable interest in extending the established methods of integrand
reduction~\cite{Ossola:2006us,Ossola:2007ax,Mastrolia:2008jb,Mastrolia:2010nb,Cullen:2011ac}
and (generalised)
unitarity~\cite{Bern:1994zx,Bern:1994cg,Bern:1997sc,Britto:2004nc,Forde:2007mi,Ellis:2007br,Giele:2008ve,Berger:2008sj} to two loops
and beyond~\cite{Mastrolia:2011pr,Kosower:2011ty,Badger:2012dp,Zhang:2012ce,Badger:2012dv,Mastrolia:2012an,Mastrolia:2012wf,
Mastrolia:2013kca,CaronHuot:2012ab,Mastrolia:2016dhn,Peraro:2016wsq,Abreu:2017idw}.
For planar $2\to 3$ scattering in Quantum Chromodynamics (QCD) this
effort has led to analytic results for the all-plus helicity two-loop
amplitude~\cite{Badger:2013gxa,Badger:2015lda,Dunbar:2016aux,Dunbar:2016cxp,Dunbar:2016gjb,Badger:2016ozq,Dunbar:2017nfy}.
The remaining helicity configurations have been obtained numerically~\cite{Badger:2017jhb,Abreu:2017hqn,Badger:2018gip,Abreu:2018jgq}.
Some groups have been able to construct solutions to the integration-by-parts reduction identities analytically
\cite{Boels:2018nrr,Chawdhry:2018awn} yet no complete amplitudes were obtained in compact form.

One of the major challenges in this program has been to understand how to efficiently build in simplifications
from integration-by-parts identities (IBPs)~\cite{Tkachov:1981wb,Chetyrkin:1981qh} that first appear
at two loops~\cite{Gluza:2010ws,Ita:2015tya,Larsen:2015ped,Georgoudis:2016wff,Kosower:2018obg,Boehm:2018fpv,Boehm:2017wjc}. 
A conventional approach to
solving this reduction problem with the Laporta algorithm~\cite{Laporta:2001dd} can be extremely computationally
intensive, especially in cases with many kinematic scales. On-going work continues to produce more and
more efficient algorithms~\cite{vonManteuffel:2012np,Smirnov:2014hma,Maierhoefer:2017hyi}.  The use
of finite field arithmetic has also been shown to provide a highly efficient method which can avoid
traditional bottlenecks~\cite{vonManteuffel:2014ixa}. It is this last approach which we build on in
this paper. It has also been demonstrated how this technique can be applied to compute scattering
amplitudes through multivariate functional reconstruction~\cite{Peraro:2016wsq}.

Another major step in the evaluation of five-point two-loop scattering amplitudes is the computation of a basis of
integral functions. Considerable progress on this front has been made recently and the analytic
evaluation of many of the two loop integrals required after reduction has been
completed with the help of differential equation methods~\cite{Kotikov:1990kg,Gehrmann:1999as,Henn:2013pwa,Papadopoulos:2014lla,vonManteuffel:2014qoa,Gehrmann:2015bfy,Papadopoulos:2015jft,
Zeng:2017ipr,Chicherin:2017dob,Gehrmann:2018yef,Abreu:2018rcw,Chicherin:2018mue}.

Analytic results can offer many benefits over numerical algorithms. The one-loop amplitudes for
five-gluon scattering, first derived in 1993 by Bern, Dixon and Kosower~\cite{Bern:1993mq}, are
strikingly simple. One immediate consequence of this is that amplitudes are fast and stable to
evaluate numerically and well suited for Monte Carlo integration. Analytic results also give us more
insight into the structure of on-shell amplitudes in gauge theory. Simplicity in maximally
super-symmetric Yang-Mills theory has enabled huge leaps into the structure of perturbative
amplitudes based on constraints from universal behaviour in physical limits~\cite{Caron-Huot:2018dsv,Dixon:2016nkn,Caron-Huot:2016owq,Dixon:2015iva}. While in QCD these
constraints are not quite enough to fix the amplitudes (such techniques have been applied in the
computation of the QCD soft anomalous dimension~\cite{Almelid:2017qju}), it would be an extremely powerful tool if the
function space of multi-loop amplitudes could be better understood in general gauge theories.

In this paper we present new, analytic results for the scattering of five gluons in pure Yang-Mills at
two loops in which one gluon has negative helicity and the remaining gluons have positive
helicities. We employ finite field numerics to a combined system of integrand reduction,
integration-by-parts identities and expansion into a basis of pentagon functions. After multiple
evaluations we were able to reconstruct the analytic form of the amplitude.

We outline our conventions and notation in Section~\ref{sec:conventions}. We then describe the
numerical algorithm used to map the coefficients of a pentagon function basis for the finite
remainder of the two-loop amplitude from generalised unitarity cuts with six-dimensional tree
amplitudes in Section~\ref{sec:setup}. This numerical algorithm is then sampled using finite field arithmetic and the rational
coefficients of the polylogarithmic pentagon functions are reconstructed as functions of momentum
twistor variables. We present our results in Section~\ref{sec:results} before drawing some brief
conclusions.

\section{Conventions and notation \label{sec:conventions}}

We compute the leading colour contribution to five-gluon scattering in pure Yang-Mills in the fundamental trace basis:
\begin{align}
  \mathcal{A}^{(L)}(1_g,2_g,3_g,4_g,5_g) = n^L g_s^3  \sum_{\sigma \in S_5/Z_5} & \tr \left(
  T^{a_{\sigma(1)}} T^{a_{\sigma(2)}} T^{a_{\sigma(3)}} T^{a_{\sigma(4)}} T^{a_{\sigma(5)}} \right) \nonumber\\
& \times
  A^{(L)}\left(\sigma(1)_g,\sigma(2)_g,\sigma(3)_g,\sigma(4)_g,\sigma(5)_g \right)
\end{align}
where $L$ is the number of loops and we have extracted a normalisation defined by,
\begin{equation}
  n= m_\eps N_c \alpha_s/(4\pi),\quad \alpha_s = g_s^2/(4\pi),\quad m_\eps=i (4\pi)^{\eps} e^{-\eps\gamma_E}.
\end{equation}
We further expand the amplitudes around $d_s=2$ where $d_s = g^\mu{}_\mu$ is the spin dimension,
\begin{subequations}
\begin{align}
  A^{(1)}\left(1^-_g,2^+_g,3^+_g,4^+_g,5^+_g \right) = & (d_s-2) A^{(1),[1]}\left(1^-_g,2^+_g,3^+_g,4^+_g,5^+_g \right), \\
  A^{(2)}\left(1^-_g,2^+_g,3^+_g,4^+_g,5^+_g \right) = & \sum_{i=0}^{2} (d_s-2)^i A^{(2),[i]}\left(1^-_g,2^+_g,3^+_g,4^+_g,5^+_g \right).
\end{align}
\end{subequations}
This is useful since the $d_s=2$ limit behaves like a supersymmetric amplitude where additional cancellations and simplifications can be seen.
In the case of the single-minus helicity configuration, it was already observed that
$A^{(2),[0]}\left(1^-_g,2^+_g,3^+_g,4^+_g,5^+_g \right) = \mathcal{O(\eps)}$~\cite{Badger:2017jhb}.

Since the tree-level helicity amplitude is zero, the universal infrared (IR) poles take a very simple
form~\cite{Catani:1998bh,Becher:2009qa,Becher:2009cu,Gardi:2009qi},
\begin{subequations}
  \begin{align}
    A^{(1),[1]}\left(1^-_g,2^+_g,3^+_g,4^+_g,5^+_g \right) = &\ F^{(1),[1]}\left(1^-_g,2^+_g,3^+_g,4^+_g,5^+_g \right) + \mathcal{O(\eps)},\\
  A^{(2),[1]}\left(1^-_g,2^+_g,3^+_g,4^+_g,5^+_g \right) =
  & \left[-\frac{r_\Gamma }{\eps^2}\sum_{i=1}^{5} \left( \frac{\mu_{R}^2 e^{\gamma_E}}{-s_{i,i+1}} \right)^\eps \right]  A^{(1),[1]}\left(1^-_g,2^+_g,3^+_g,4^+_g,5^+_g \right)
  \label{eq:irpoles}
  \nn
  & + F^{(2),[1]}\left(1^-_g,2^+_g,3^+_g,4^+_g,5^+_g \right) + \mathcal{O(\eps)},\\
  A^{(2),[2]}\left(1^-_g,2^+_g,3^+_g,4^+_g,5^+_g \right) = &\ F^{(2),[2]}\left(1^-_g,2^+_g,3^+_g,4^+_g,5^+_g \right) + \mathcal{O(\eps)},
\end{align}
\end{subequations}
where
\begin{equation}
r_\Gamma = \frac{\Gamma^2(1-\eps)\Gamma(1+\eps)}{\Gamma(1-2\eps)}.
\end{equation}
In this paper we will present a direct computation of the finite remainder $F$.

\section{Computational setup \label{sec:setup}}

The kinematic parts of the amplitude are written using a momentum
twistor~\cite{Hodges:2009hk} parametrisation, as described in previous
works~\cite{Badger:2013gxa,Badger:2017jhb}.
We decompose the amplitude into an integrand
basis, using the method of integrand reduction via generalised
unitarity.  We then reduce the amplitude to master integrals by
solving IBPs.  The master integrals
are in turn expressed as combinations of known pentagon
functions, using the expressions computed in
reference~\cite{Gehrmann:2018yef}.

The algorithm is implemented numerically over finite fields.
The Laurent expansion in $\epsilon$ of the results is obtained
by performing a full reconstruction of its dependence on the
dimensional regulator $\epsilon$, for fixed numerical values of the
kinematic variables.  The Laurent expansion of the reconstructed
function of $\epsilon$ thus provides a numerical evaluation of the
$\epsilon$-expansion of the final result.  Finally, the full
dependence of the expanded result on the kinematic variables is
reconstructed from multiple numerical evaluations, using a modified
version of the multi-variate reconstruction techniques presented in
reference~\cite{Peraro:2016wsq}.

In this section we provide more details on our computational setup
and the various steps of the calculation outlined above.

\subsection{Integrand parametrisation \label{sec:integrandreduction}}

\begin{figure}
	\begin{subfigure}{.33\textwidth}
		\centering
		\includegraphics[totalheight=3.2cm]{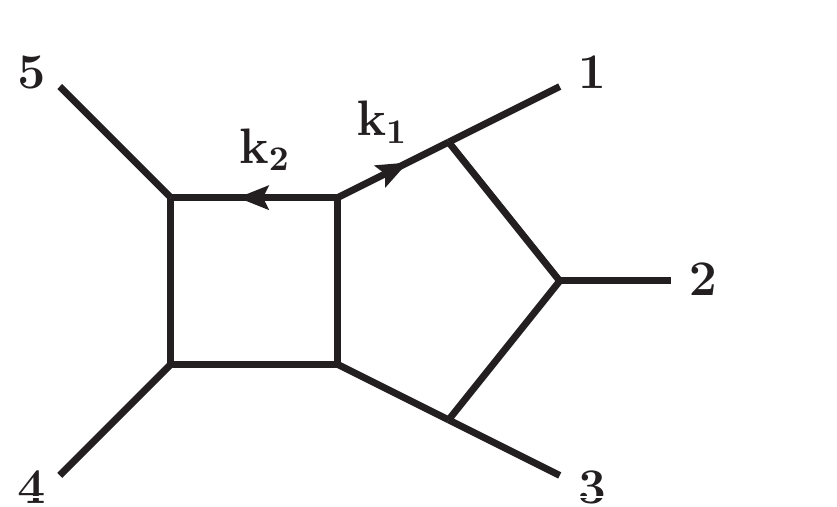}
		\caption{{\footnotesize Pentabox: $G_{11111111a_{9}a_{10}a_{11}}$}}
		\label{fig:pentabox}
	\end{subfigure}%
	\begin{subfigure}{.33\textwidth}
		\centering
		\includegraphics[totalheight=3.4cm]{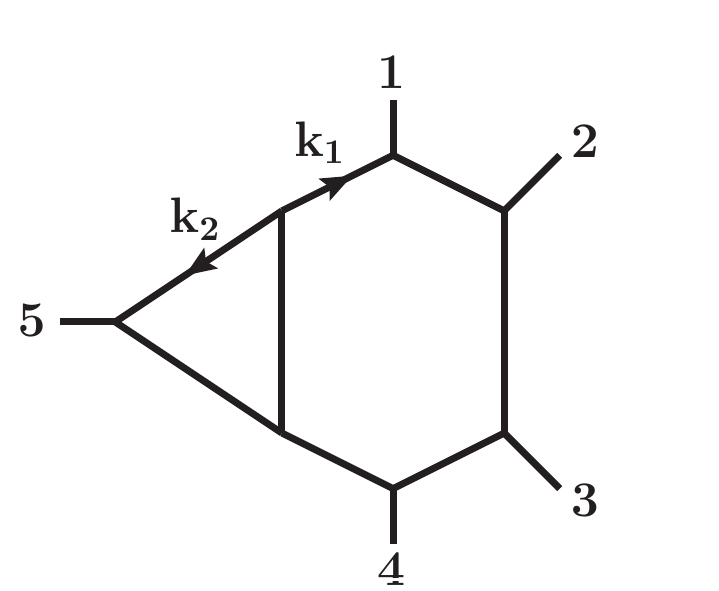}
		\caption{{\footnotesize Hexatriangle: $G_{111111a_{7}11a_{10}a_{11}}$}}
		\label{fig:hexatriangle}
	\end{subfigure}%
	\begin{subfigure}{.34\textwidth}
		\centering
		\includegraphics[totalheight=3.3cm]{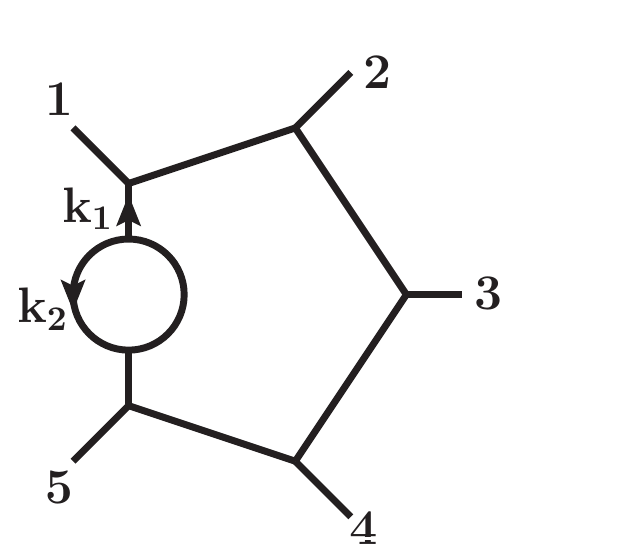}
		\caption{{\footnotesize Heptabubble: $G_{21111a_{6}a_{7}11a_{10}a_{11}}$}}
		\label{fig:heptabubble}
	\end{subfigure}
	\caption{Two-loop five-point master topologies. All external momenta are considered out-going, arrows indicate loop momenta directions.}
	\label{fig:mastertopos}
\end{figure}

We define an integral family by a complete, minimal set of propagators and irreducible scalar products (ISPs):
\begin{align}
  G_{a_1 a_2 a_3 a_4 a_5 a_6 a_7 a_8 a_9 a_{10} a_{11}} &=
  \int \frac{d^d k_1}{i \pi^{d/2} e^{-\eps \gamma_E}} \frac{d^d k_2}{i \pi^{d/2} e^{-\eps \gamma_E}} \nonumber \\
  &\times \frac{1}{k_1^{2a_1}} \frac{1}{(k_1-p_1)^{2a_2}} \frac{1}{(k_1-p_1-p_2)^{2a_3}} \frac{1}{(k_1+p_4+p_5)^{2a_4}} \nonumber \\
  &\times \frac{1}{k_2^{2a_5}} \frac{1}{(k_2-p_5)^{2a_6}} \frac{1}{(k_2-p_4-p_5)^{2a_7}} \frac{1}{(k_1+k_2)^{2a_8}} \nonumber \\
  &\times \frac{1}{(k_1+p_5)^{2a_9}} \frac{1}{(k_2+p_1)^{2a_{10}}} \frac{1}{(k_2+p_1+p_2)^{2a_{11}}},
\end{align}
where the exponents, $a_i$, are integers and $d=4-2\eps$. The three master topologies, shown in Figure~\ref{fig:mastertopos}, are
\begin{alignat}{2}
  &\text{Pentabox:}& &G_{1 1 1 1 1 1 1 1 a_9 a_{10} a_{11}}\label{eq:pentabox} \\
  &\text{Hexatriangle:}& &G_{1 1 1 1 1 1 a_7 1 1 a_{10} a_{11}}\label{eq:hexatriangle} \\
  &\text{Heptabubble:}\quad& &G_{2 1 1 1 1 a_6 a_7 1 1 a_{10} a_{11}}\label{eq:heptabubble}
\end{alignat}
while propagators with unspecified exponents, $a_j$, correspond to ISPs (i.e.\ $a_j\leq 0$).

\begin{figure}
	\begin{subfigure}{.45\textwidth}
		\centering
		\includegraphics[width=6.9cm,keepaspectratio]{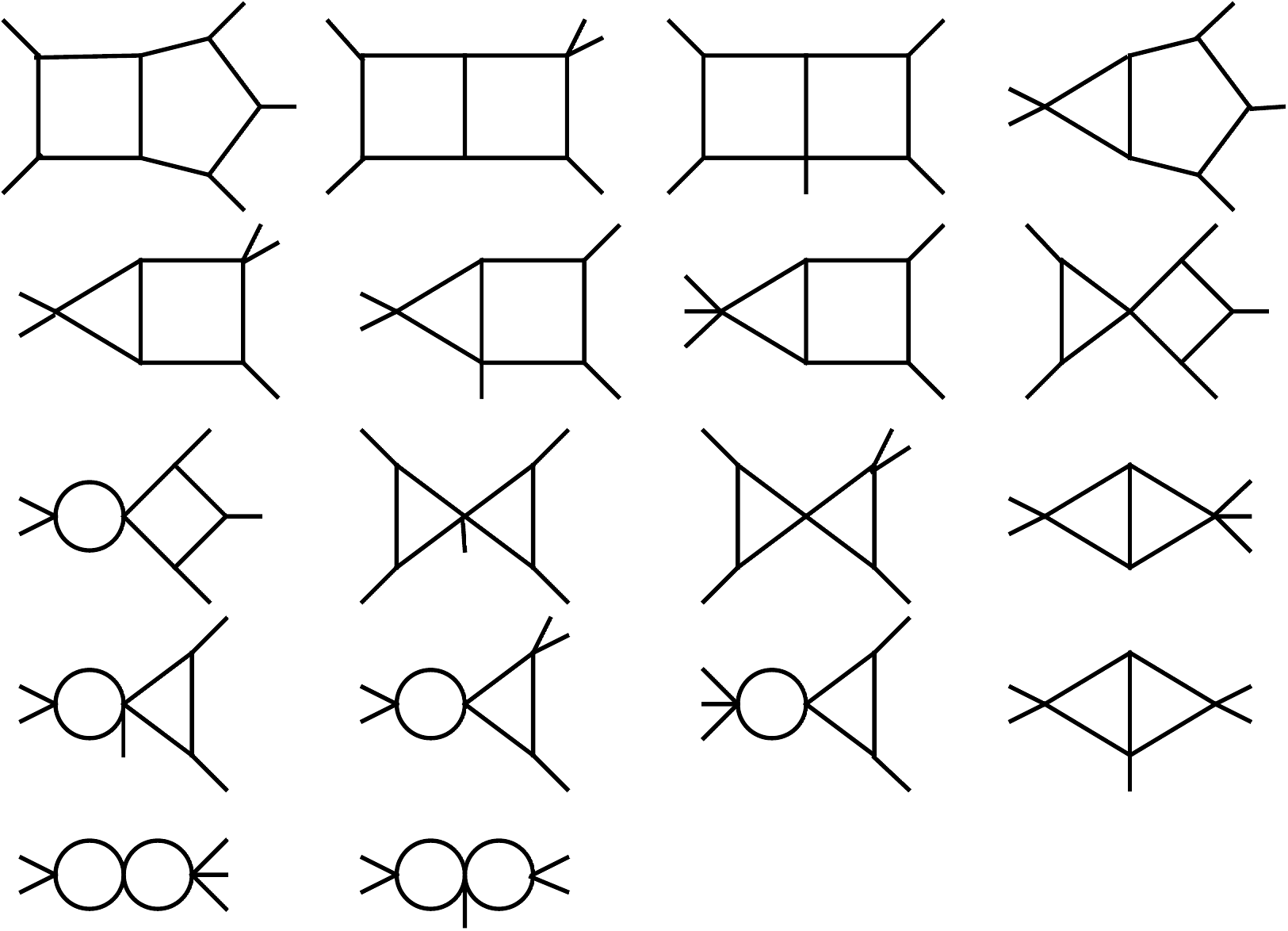}
		\caption{Topologies associated with the pentabox master topology (top left), see~\eqref{eq:pentabox}.}
		\label{fig:pentaboxsector}
	\end{subfigure}
	\begin{subfigure}{.45\textwidth}
		\centering
		\includegraphics[width=6.9cm,keepaspectratio]{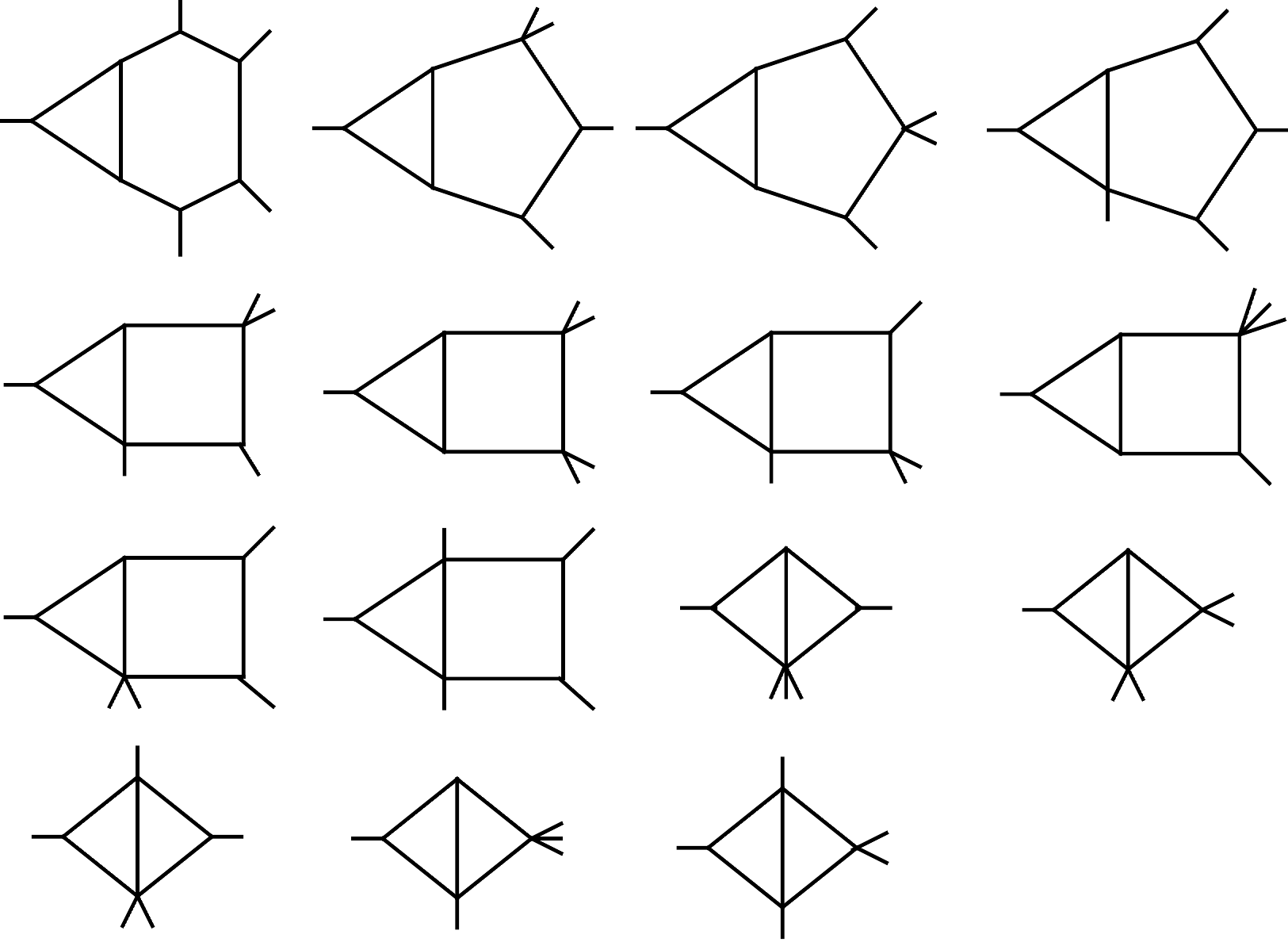}
		\caption{Topologies associated with the hexatriangle master topology (top left), see~\eqref{eq:hexatriangle}.}
		\label{fig:hexatrianglesector}
	\end{subfigure}
	\begin{subfigure}{.45\textwidth}
		\centering
		\includegraphics[width=6.9cm,keepaspectratio]{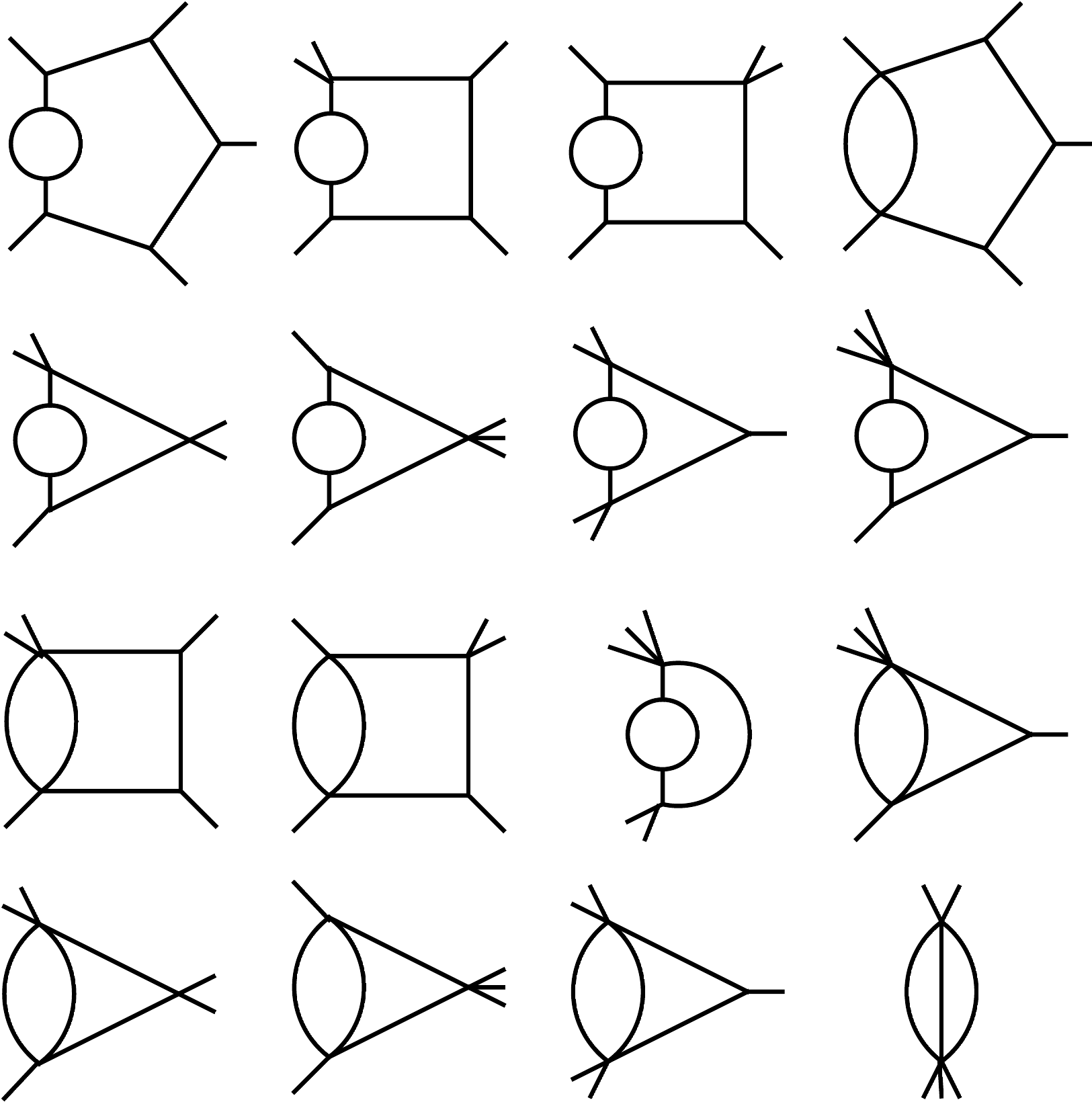}
		\caption{Topologies associated with the heptabubble master topology (top left), see~\eqref{eq:heptabubble}.}
		\label{fig:heptabubblesector}
	\end{subfigure}\hspace{1.5cm}
	\begin{subfigure}{.45\textwidth}
		\centering
		\includegraphics[width=6.9cm,keepaspectratio]{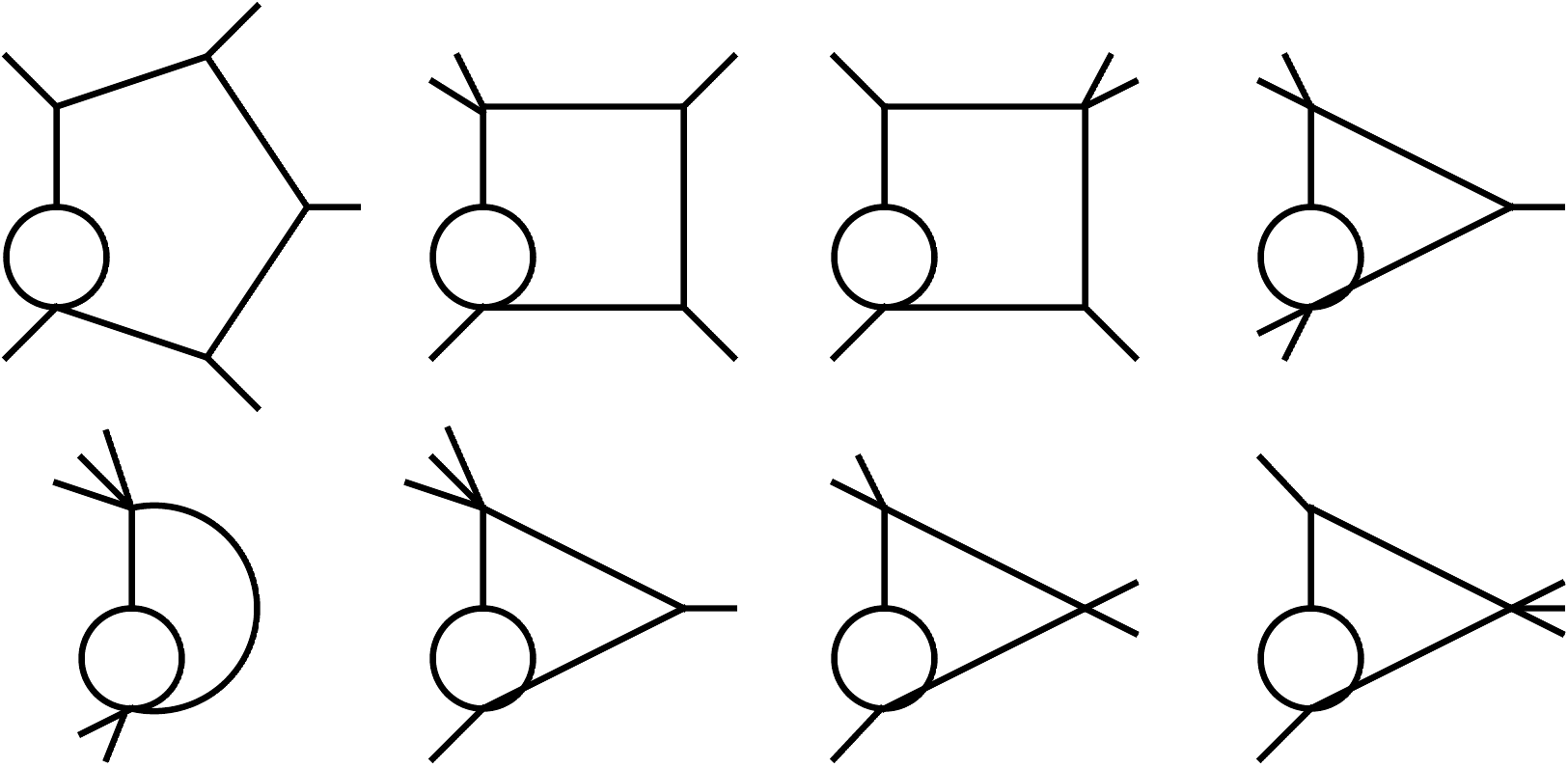}
		\caption{Topologies with divergent cuts that must be computed simultaneously with sub-topologies of the heptabubble master topology in Figure~\ref{fig:heptabubblesector}.}
		\label{fig:divergenttopos}
	\end{subfigure}
	\caption{All distinct two-loop five-point topologies.}
	\label{fig:alltopos}
\end{figure}

All lower point topologies are obtained by systematically pinching the propagators of the master
topologies. Topologies with scaleless integrals are discarded since we work in dimensional
regularisation. Pinching of propagators from different master topologies can lead to the same
sub-topology. This happens in particular when all five cyclic permutations of the external momenta
are included. In these cases the assignment to a master topology is not unique. The full set of 57
distinct topologies with a specific choice of master topology assignment is shown in Figure
\ref{fig:alltopos}.

We parametrise the integrand numerators by writing the most general polynomials in the ISPs subject
to a power counting constraint from renormalizability considerations. As an example, the pentabox of
Figure~\ref{fig:pentabox} has the numerator parametrisation,
\begin{align}
\Delta \left( \parbox{1.7cm}{ \includegraphics[width=2cm]{figures/pentabox}} \right) =
 \sum c_{(1,1,1,1,1,1,1,1,a_9,a_{10},a_{11})} (k_1+p_5)^{-2 a_{9}} (k_2+p_1)^{-2 a_{10}} (k_2+p_1+p_2)^{-2 a_{11}}
 \label{eq:pentaboxnumerator}
\end{align}
where the sum is truncated by the constraints on the exponents:
\begin{alignat}{2}
-5 & \leq a_9 &&\leq 0,  \\
-4 & \leq a_{10}+a_{11} &&\leq 0, \\
-7 & \leq a_9+a_{10}+a_{10} &&\leq 0.
\end{alignat}
Each topology has $11-n$ ISPs where $n$ is the number of distinct propagators. The five cyclic permutations of the external legs give a total of 425 irreducible numerators.

Integrand representations of the form~\eqref{eq:pentaboxnumerator} are less compact than
representations making use of, for example, local integrands, spurious integrands, and
extra-dimensional
ISPs~\cite{ArkaniHamed:2010kv,ArkaniHamed:2010gh,Badger:2013gxa,Badger:2015lda,Badger:2016ozq,Badger:2017jhb,Bourjaily:2017wjl}.  However,
in our set-up the integrand is only sampled numerically and not analytically reconstructed. Simplification at the integrand level is therefore not a priority.
Because our final integrated amplitude does not depend on the choice of integrand
parametrisation, we have chosen a form which is directly compatible with IBPs, rather than one yielding a compact integrand representation.
On the other hand there is potential for considerable improvements in the efficiency of the
algorithm if a simpler integrand form could be identified.

We take a top down approach to solving the complete system of integrands which, apart from the basis
choice described above, is identical to the approach taken in Ref.~\cite{Badger:2017jhb}.
The tree amplitudes used to compute the generalised unitarity cuts are evaluated by contracting Berends-Giele currents~\cite{Berends:1987me} as
described in~\cite{Peraro:2016wsq} and
the six-dimensional spinor-helicity formalism~\cite{Cheung:2009dc}.
Eight topologies, shown in Figure~\ref{fig:divergenttopos}, have divergent cuts and their integrand coefficients are determined
simultaneously with sub-topologies in the heptabubble group, see Figure~\ref{fig:heptabubblesector}.

The numerical sampling of the integrand can show quickly which coefficients vanish and hence what integrals
require further reduction using IBPs. The number of non-vanishing coefficients at the integrand level split
into the components of $d_s=2, (d_s-2), (d_s-2)^2$ are $4387,14545,4420$ respectively. We find the
maximum rank to be 5 for genuine two-loop topologies and rank 6 for a few integrals in the $(d_s-2)^2$
component of the amplitude that can be written as (1-loop)${}^2$ integrals, see Figure~\ref{fig:pentaboxsector}.

At the end of the integrand reduction stage, the colour ordered amplitude can be written as
\begin{align} \label{eq:A2}
  A^{(2)}\left( 1,2,3,4,5 \right) = \sum_{\mathbf{a}} \cintegrand_{\mathbf{a}}\ G_{\mathbf{a}},
\end{align}
where we sum over the tuples $ \mathbf{a} = (a_1,a_2,a_3,a_4,a_5,a_6,a_7,a_8,a_9,a_{10},a_{11})$. The coefficients $\cintegrand_{\mathbf{a}}$ are rational functions in the momentum
twistor variables only.

\subsection{Integration-by-parts identities \label{sec:ibps}}

Each integral appearing in~\eqref{eq:A2} is reduced to a set of master
integrals $J_k$,
\begin{equation}
  G_{\mathbf{a}} = \sum_{k} \cibp_{\mathbf{a} k} J_k,
  \label{eq:ibpreduction}
\end{equation}
where the sum runs over 155 master integrals (remembering that we
include the 5 cyclic permutations of the integral family $G$).  The
reduction is obtained by solving a traditional Laporta system of
IBP equations~\cite{Laporta:2001dd}.  The system is generated
in \textsc{Mathematica} with the help of \textsc{LiteRed}~\cite{Lee:2012cn}, and solved over finite fields with a custom general-purpose
linear solver for sparse systems of equations.  The master integrals
are chosen to be the uniform weight functions identified by Gehrmann,
Henn and Lo Presti~\cite{Gehrmann:2018yef}. The $\cibp_{\mathbf{a} k}$ are rational
functions in the momentum twistor variables and the dimensional
regularisation parameter $\eps$.

\subsection{Map to pentagon functions \label{sec:pfunc}}

Our next step is to expand the master integrals into a basis of pentagon functions defined by
Gehrmann, Henn and Lo Presti. These functions can be written in terms of Goncharov Polylogarithms.
We take the results of expanding the master integrals in $\eps$ from reference~\cite{Gehrmann:2018yef},
\begin{equation}
  J_k = \sum_{x=0}^4 \sum_l \cpfunc_{kl;x}\, \eps^x\, m_{l;x}(f) + \mathcal{O}(\eps^5),
  \label{eq:pfuncbasis}
\end{equation}
where $m_{l;x}(f)$ are monomials in the pentagon functions.

The amplitude can thus be written as a combination of pentagon
functions
\begin{align}
  A^{(2)}\left( 1,2,3,4,5 \right) = \sum_{l,x} \camp_{l;x} m_{l;x}(f) + \mathcal{O}(\eps),
\end{align}
where the coefficients are defined through matrix multiplication, from the three reduction steps
\begin{equation}
  \camp_{l;x} = \sum_{\mathbf{a},k} \sum_{x=0}^4 \cintegrand_{\mathbf{a}}\, \cibp_{\mathbf{a} k}\, \cpfunc_{kl;x}\, \eps^x.
\end{equation}
We recall that, in the previous equation, there is also an implicit
dependence on $\epsilon$ coming from the $\cibp_{\mathbf{a} k}$, which were
defined in~\eqref{eq:ibpreduction} to be the full coefficients of the
IBP reduction.  Hence, the coefficients $\camp_{l;x}$ are rational
functions of $\epsilon$ which need to be expanded, as we will explain
in the next subsection.

The final step of the algorithm is to perform the same decomposition for the universal IR poles
in~\eqref{eq:irpoles}. For this we need the one-loop master integrals expanded up to weight four and
written in the same alphabet as the two-loop integrals. These results were obtained directly
from the differential equations in a canonical basis.\footnote{We are very grateful to Adriano Lo Presti for
assistance in setting up the differential equations used in~\cite{Gehrmann:2015bfy}.}  We then write the
poles analytically as
\begin{align}
  \left[-\frac{r_\Gamma }{\eps^2}\sum_{j=1}^{5} \left( \frac{\mu_{R}^2 e^{\gamma_E}}{-s_{j,j+1}}
  \right)^\eps \right]  A^{(1),[i]}\left(1^-_g,2^+_g,3^+_g,4^+_g,5^+_g \right) = \sum_{l,x}
  \cir_{l;x} m_{l;x}(f) + \mathcal{O}(\eps).
\end{align}
Our numerical algorithm can then compute the difference:
\begin{equation}
  \label{eq:cF}
  \cfin_{l;x} = \camp_{l;x} - \cir_{l;x}
\end{equation}
which we will expand in $\eps$ to find the finite remainder. At this point we have constructed a numerical algorithm which combines integrand reduction, IBP
reduction and expansion of the master integrals into a basis of polylogarithms. This algorithm can be used to compute the finite remainder of the two-loop amplitude through evaluations of generalised
unitarity cuts over finite fields.

\subsection{Laurent expansion  \label{sec:laurent}}

In the previous subsections we described a numerical calculation over
finite fields of the coefficients $\cfin_{l;x}$. They are used in order to write the finite
remainder, $F^{(2),[i]},$ of the amplitude in terms of known pentagon
functions.  The coefficients, computed as described above, are rational
functions of $\eps$.  However, because the calculation uses the
expansion in~\eqref{eq:pfuncbasis} for the master integrals in terms
of pentagon functions, it is only valid up to $\mathcal{O}(\eps)$.
Here we are interested in the finite part of the Laurent expansion in
$\epsilon$.

As mentioned before, in order to obtain this Laurent expansion, we
first perform a full reconstruction of the functions $\cfin_{l;x}$
in $\epsilon$, for numerical values over finite fields of the momentum twistor variables.  The reconstructed function can thus be
expanded in $\eps$ up to the desired order.  This yields a
decomposition of the form
\begin{equation}
  \label{eq:cFepsexp}
  \cfin_{l;x}  = \sum_{y=-4}^0\, \cfin_{l;x,y} \eps^y + \mathcal{O}(\eps),
\end{equation}
where we are interested in the finite parts $\cfin_{l;x,0}$, while
$\cfin_{l;x,y}=0$ for $y<0$.  The finite remainder is therefore
\begin{align}\label{eq:F2}
  F^{(2),[1]}\left(1^-_g,2^+_g,3^+_g,4^+_g,5^+_g \right) = \sum_{l} \sum_{x=0}^4 \cfin_{l;x,0}\, m_{l;x}(f) + \mathcal{O}(\eps),
\end{align}
with $\cfin_{l;x,0}$ defined by the Laurent expansion
in~\eqref{eq:cFepsexp}.

The algorithm described above numerically computes the
coefficients $\cfin_{l;x,0}$ of the finite remainder of the
amplitude over finite fields, for any numerical value of the kinematic
invariants represented by the momentum twistor variables.  Full
analytic formulas for the coefficients $\cfin_{l;x,0}$, as
rational functions of the momentum twistor variables, are
reconstructed from multiple numerical evaluations.  For this purpose,
we use a slightly improved version of the multivariate reconstruction
techniques presented in reference~\cite{Peraro:2016wsq}.

In the next section we give a compact form of this result, obtained
from the one in terms of momentum twistor variables, after converting it into
spinor products and momentum invariants via some
additional algebraic manipulations.

\section{Results \label{sec:results}}

We present a compact form of the amplitude by making use of the symmetry $(1,2,3,4,5)\rightarrow(1,5,4,3,2)$
and extracting an overall phase written in terms of spinor products,
\begin{equation}
F^{(L),[i]}\left(1^-_g,2^+_g,3^+_g,4^+_g,5^+_g \right) = \frac{\spB25^2}{\spB12\spA23\spA34\spA45\spB51}
	\left( F_{\rm sym}^{(L),[i]}(1,2,3,4,5) + F_{\rm sym}^{(L),[i]}(1,5,4,3,2) \right),
\end{equation}
where $L$ labels the loop order and $i$ labels the component in the expansion around $d_s=2$. The known result at one-loop can be written as:
\begin{equation}
F_{\rm sym}^{(1),[1]}(1,2,3,4,5) = \frac{\trp(2315)^2 \trp(1243)}{3 s_{25}^2s_{23}s_{34}s_{15}} - \frac{\trp(2543)}{6 s_{34}},
\end{equation}
where $\trp(ijkl) = \frac{1}{2}\tr( (1+\gamma_5)\slashed{p}_i\slashed{p}_j\slashed{p}_k\slashed{p}_l )$ and $s_{ij} = (p_i+p_j)^2$.

The finite parts of the two-loop amplitude can be written compactly in terms of weight two functions, just as at one-loop.
We therefore follow the same strategy as at one-loop to find a basis of integral functions free of large cancellations due to spurious
singularities. We find that a convenient basis for the $d_s-2$ component of the amplitude is
\begin{align}
F_{\rm sym}^{(2),[1]}(1,2,3,4,5)  =
& \quad c_{51}^{(2)} \fbox2 (s_{23},s_{34},s_{15})
  + c_{51}^{(1)} \fbox1 (s_{23},s_{34},s_{15})
  + c_{51}^{(0)} \fbox0 (s_{23},s_{34},s_{15})  \nn
& + c_{34}^{(2)} \fbox2 (s_{12},s_{15},s_{34})
  + c_{34}^{(1)} \fbox1 (s_{12},s_{15},s_{34})
  + c_{34}^{(0)} \fbox0 (s_{12},s_{15},s_{34})  \nn
& + c_{45} \fbox0 (s_{12},s_{23},s_{45})
  + c_{34;51} \lh_1(s_{34},s_{15})
  + c_{51;23} \lh_1(s_{15},s_{23})
  + c_{\rm rat},
\end{align}
and
\begin{align}
F_{\rm sym}^{(2),[2]}(1,2,3,4,5)  =
& \quad d_{34}^{(3)} \fbox3 (s_{12},s_{15},s_{34})
  + d_{34}^{(2)} \fbox2 (s_{12},s_{15},s_{34})   \nn
& + d^{(3)}_{34;51} \lh_3(s_{34},s_{15})
  + d^{(2)}_{34;51} \lh_2(s_{34},s_{15}) \nn
& + d^{(3)}_{51;23} \lh_3(s_{15},s_{23})
  + d^{(2)}_{51;23} \lh_2(s_{15},s_{23})
  + d_{\rm rat},
\end{align}
for the $(d_s-2)^2$ amplitude.

The integral functions are written in terms of simple logarithms and di-logarithms. All weight one functions
appear as logarithms of ratios of kinematic invariants,
\begin{align}
L_k(s,t) =  & \frac{\log(t/s)}{(s-t)^k},
\end{align}
where the singular behaviour is removed by defining,
\begin{subequations}
\begin{align}
\lh_{0}(s,t) = & L_0(s,t), \\
\lh_{1}(s,t) = & L_1(s,t), \\
\lh_{2}(s,t) = & L_2(s,t) + \frac{1}{2(s-t)}\left(\frac{1}{s}+\frac{1}{t}\right),  \\
\lh_{3}(s,t) = & L_3(s,t) + \frac{1}{2(s-t)^2}\left(\frac{1}{s}+\frac{1}{t}\right).
\end{align}
\end{subequations}
At weight two all functions can be written in terms of the six-dimensional box function,
\begin{subequations}
\begin{align}
\fbox{-1}(s,t,m^2) = & \li2\left(1-\frac{s}{m^2}\right) + \li2\left(1-\frac{t}{m^2}\right) + \log\left(\frac{s}{m^2}\right) + \log\left(\frac{t}{m^2}\right) - \frac{\pi^2}{6}, \\
\fbox{0}(s,t,m^2) = & \frac{1}{u(s,t,m^2)} \fbox{-1}(s,t,m^2), \\
\fbox{1}(s,t,m^2) = & \frac{1}{u(s,t,m^2)} \left[ \fbox{0}(s,t,m^2) + \lh_1(s,m^2) + \lh_1(m^2,t) \right], \\
\fbox{2}(s,t,m^2) = & \frac{1}{u(s,t,m^2)} \Bigg[  \fbox{1}(s,t,m^2)
                                              + \frac{s-m^2}{2t}\lh_2(s,m^2)
                                              + \frac{m^2-t}{2s}\lh_2(m^2,t) \nonumber\\&
                       - \left(\frac{1}{s}+\frac{1}{t}\right)\frac{1}{4m^2} \big], \\
\fbox{3}(s,t,m^2) = & \frac{1}{u(s,t,m^2)} \big[  \fbox{2}(s,t,m^2)
                                              - \frac{(s-m^2)^2}{6t^2}\lh_3(s,m^2)
                                              - \frac{(m^2-t)^2}{6s^2}\lh_3(m^2,t) \nonumber\\&
                     - \left(\frac{1}{s}+\frac{1}{t}\right)\frac{1}{6m^4} \Bigg],
\end{align}
\end{subequations}
where $u(s,t,m^2) = m^2-s-t$.

These functions serve the same purpose as the $Ls$ and $L$ functions introduced by Bern, Dixon, and Kosower in~\cite{Bern:1993mq,Bern:1994fz}. The $\lh_{i}(s,t)$
are finite as $s\to t$ and the $\fbox{i}(s,t,m^2)$ are finite as $s\to -t+m^2$. The definitions have been changed very slightly with
respect to the $Ls$ and $L$ functions since the singularities from the box functions at $m^2-s-t$ have been removed without
introducing additional singularities in $s-m^2$ or $t-m^2$.


For the $(d_s-2)$ amplitude the coefficients are:

\begin{subequations}
\begin{align}
c_{51}^{(2)} = & \frac{5 s_{23} s_{34} \trp(1234)^2 \trp(1542)^2}{s_{12} s_{15} \trp(2543)^2}, \\
c_{51}^{(1)} = & -\frac{\trp(1234)^2 \trp(1534) \trp(2453)^2}{6 s_{12} s_{34} s_{35} \trp(2543)^2}, \\
c_{51}^{(0)} = & \frac{s_{15} s_{45} \trp(1234)}{3 \trp(2543)}
                -\frac{s_{15} s_{24} s_{45} \trp(1234)^2}{6 s_{12} \trp(2543)^2}
                - \frac{\trp(1234)^2 \trp(1542)}{6 \trp(2543)^2} \nn
               & - \frac{s_{23} s_{24} \trp(1234) \trp(1543)}{6 \trp(2543)^2}
                - \frac{s_{12} s_{23} s_{34} \trp(1542) \trp(1543)}{3 s_{15} \trp(2543)^2} \nn
               & + \frac{2 s_{23} \trp(1234) \trp(1543)^2}{3 s_{15} \trp(2543)^2}
                - \frac{s_{24} \trp(1234) \trp(1543)^2}{6 s_{15} \trp(2543)^2} \nn
              &  - \frac{s_{24} \trp(1234)^2 \trp(1543)^2}{6 s_{12} s_{15} s_{34} \trp(2543)^2}
                - \frac{s_{24} \trp(1235) \trp(1243) \trp(1543)^2}{6 s_{12} s_{15} s_{34} \trp(2543)^2} \nn
              &  - \frac{\trp(1234) \trp(1543) \trp(2453)}{2 \trp(2543)^2},
\end{align}
\end{subequations}

\begin{subequations}
\begin{align}
c_{34}^{(2)} = & \frac{5}{2} s_{12}^2 s_{15}^2, \\
c_{34}^{(1)} = & \frac{\trp(1245) \trp(1534) \trp(1543)}{3 s_{15} s_{34} s_{45}}
                 +\frac{1}{12} s_{12} s_{15} s_{34}, \\
c_{34}^{(0)} = & -\frac{s_{15} s_{23} \trp(1234)}{12 \trp(2543)}
                 - \frac{\trp(1234)^2 \trp(1532)}{12 s_{12} s_{34} \trp(2543)}
                 + \frac{s_{25} \trp(1234) \trp(1543)}{12 s_{34} \trp(2543)} \nn
               & + \frac{s_{12} \trp(1532) \trp(1543)}{12 s_{15} \trp(2543)}
                 + \frac{s_{23} s_{25} \trp(1234) \trp(1543)^2}{12 s_{15} s_{34} \trp(2543)^2} \nn
               & + \frac{s_{12} \trp(1543)^2 \trp(2354)}{12 s_{15} s_{34} s_{45} \trp(2543)}
                 + \frac{s_{25} \trp(2543)}{12 s_{34}}
                 - \frac{1}{3} s_{12} s_{15},
\end{align}
\end{subequations}

\begin{equation}
c_{45} = -\frac{s_{13} s_{45} \trp(1234)^2 \trp(1534)}{6 s_{12} s_{34} \trp(2543)^2}
           +\frac{s_{23}^3 s_{34} \trp(1543)}{6 \trp(2543)^2}
           -\frac{s_{13} s_{23} \trp(1243) \trp(1543)^2}{6 s_{15} s_{34} \trp(2543)^2},
\end{equation}

\begin{align}
c_{34;51} =\ & \frac{s_{12} s_{15} s_{34} s_{45}}{6 \trp(2543)}
             - \frac{11 \trp(1234) \trp(1543)}{6 \trp(2543)}
             + \frac{\trp(1234)^2 \trp(1542) \trp(1543)}{6 s_{12} s_{34} \trp(2543)^2} \nn
            & - \frac{s_{12} \trp(1324) \trp(1543)^2}{6 s_{13} \trp(2543)^2}
             - \frac{s_{45} \trp(1234)^2 \trp(1534) \trp(2453)}{6 s_{12} s_{34} s_{35} \trp(2543)^2}, \\
c_{51;23} =\ & \frac{s_{23} s_{45} \trp(1234)^2 \trp(1534)}{6 s_{12} s_{34} \trp(2543)^2}
             + \frac{2 s_{12} s_{23} s_{34} s_{45} \trp(1543)}{3 \trp(2543)^2}  \nn
            & - \frac{5 s_{23} s_{45} \trp(1234) \trp(1543)}{2 \trp(2543)^2},  \\
c_{\rm rat} =\ & -\frac{5 \trp(1234) \trp(1543)}{4 s_{34} \trp(2543)}
        + \frac{5 s_{23} \trp(1243) \trp(1543)^2}{2 s_{15} s_{34} \trp(2543)^2}.
\end{align}

While for the $(d_s-2)^2$ amplitude the coefficients are:

\begin{subequations}
\begin{align}
d_{34}^{(3)} = & -\frac{s_{12} s_{15} \trp(2543)^2}{12 s_{34}}, \\
d_{34}^{(2)} = & \frac{1}{6} s_{12} s_{15} \trp(2543),
\end{align}
\begin{align}
d_{34;51}^{(3)} = & -\frac{1}{18} s_{15} s_{23} \trp(2543), \\
d_{34;51}^{(2)} = & -\frac{s_{15} \trp(2543)^2}{36 s_{12} s_{34}}
                    - \frac{1}{6} s_{15} \trp(2543),
\end{align}
\end{subequations}

\begin{subequations}
\begin{align}
d_{51;23}^{(3)} = & \frac{s_{15} s_{23} s_{34}^2 s_{45}^2 \trp(1245)}{18 \trp(2543)^2}, \\
d_{51;23}^{(2)} = & \frac{s_{12} s_{15} s_{23} s_{34}^2 s_{45}^2}{12 \trp(2543)^2}
                   -\frac{s_{15} s_{23} s_{34} s_{45}^2 \trp(1234)}{6 \trp(2543)^2},
\end{align}
\end{subequations}

\begin{align}
d_{\rm rat} = &   \frac{s_{34}}{72} + \frac{5 s_{45}}{36}
        - \frac{\trp(1234) \trp(1453)}{72 s_{14} \trp(2543)}
        - \frac{s_{45} \trp(1543)}{72 \trp(2543)}
        + \frac{\trp(2543)^2}{72 s_{34}^3} \nn
      & - \frac{5 \trp(1543)^2}{72 s_{34} \trp(2543)}
        - \frac{s_{12} \trp(1543)^2}{72 \trp(2543)^2}
        + \frac{s_{23} s_{34} s_{45} \trp(1543)}{18 \trp(2543)^2}.
\end{align}

These results can also be found in the ancillary file included with the arXiv version of this article.


\section{Conclusions \label{sec:conclusions}}

In this article we have presented a new analytic result for a two-loop five-gluon scattering
amplitude in QCD. We were able to find a compact representation for the finite remainder of the
single-minus helicity configuration after removing the universal infrared poles.
We set up a complete tool-chain from generalised unitarity cuts to the coefficients of a basis of pentagon functions
for the finite remainder. This numerical algorithm was then evaluated multiple times with finite field arithmetic and the analytic
result reconstructed, avoiding the usual large intermediate expressions.

This single-minus amplitude has turned out to be significantly more difficult to compute than
the highly symmetric all-plus helicity amplitude that has been known for some
time~\cite{Badger:2013gxa,Gehrmann:2015bfy}. At the level of the master integrals the single-minus
amplitude was of similar complexity to the maximal-helicity-violating (MHV) configurations. However, after removing the
contribution from the universal poles, the finite remainder was simple and contained only up to weight two
polylogarithms. This makes the final answer simpler than the more general MHV case which will have
up to weight four polylogarithms.

Nevertheless, the techniques presented here are not helicity dependent so we hope to find
applications to the remaining independent planar helicity amplitudes in the near future. The last
few months have also seen progress on non-planar
integrals~\cite{Chicherin:2017dob,Abreu:2018rcw,Chicherin:2018mue} for five-point scattering which
is encouraging for applications to non-planar amplitudes.

\begin{acknowledgments}
  We are extremely grateful to Johannes Henn, Thomas Gehrmann, Adriano Lo Presti, Fabrizio Caola and Calum Milloy
  for useful discussions. SB is supported by an STFC Rutherford Fellowship ST/L004925/1 and CBH and HBH are supported by Rutherford
  Grant ST/M004104/1. This project has received funding from the European Union's Horizon 2020
  research and innovation programme under grant agreement No 772099.  This project has received funding from the European Union’s Horizon 2020 research and innovation programme under the Marie Skłodowska-Curie grant agreement 746223.
\end{acknowledgments}

\appendix

\bibliographystyle{JHEP}
\bibliography{2l5gmpppp}

\providecommand{\href}[2]{#2}\begingroup\raggedright\begin{thebibliography}{10}

\bibitem{Bendavid:2018nar}
J.~R. Andersen et~al., \emph{{Les Houches 2017: Physics at TeV Colliders
  Standard Model Working Group Report}},  in \emph{{10th Les Houches Workshop
  on Physics at TeV Colliders (PhysTeV 2017) Les Houches, France, June 5-23,
  2017}}, 2018.
\newblock \href{http://arxiv.org/abs/1803.07977}{{\tt 1803.07977}}.

\bibitem{Ossola:2006us}
G.~Ossola, C.~G. Papadopoulos and R.~Pittau, \emph{{Reducing full one-loop
  amplitudes to scalar integrals at the integrand level}},
  \href{http://dx.doi.org/10.1016/j.nuclphysb.2006.11.012}{\emph{Nucl. Phys.}
  {\bf B763} (2007) 147--169}, [\href{http://arxiv.org/abs/hep-ph/0609007}{{\tt
  hep-ph/0609007}}].

\bibitem{Ossola:2007ax}
G.~Ossola, C.~G. Papadopoulos and R.~Pittau, \emph{{CutTools: A Program
  implementing the OPP reduction method to compute one-loop amplitudes}},
  \href{http://dx.doi.org/10.1088/1126-6708/2008/03/042}{\emph{JHEP} {\bf 03}
  (2008) 042}, [\href{http://arxiv.org/abs/0711.3596}{{\tt 0711.3596}}].

\bibitem{Mastrolia:2008jb}
P.~Mastrolia, G.~Ossola, C.~G. Papadopoulos and R.~Pittau, \emph{{Optimizing
  the Reduction of One-Loop Amplitudes}},
  \href{http://dx.doi.org/10.1088/1126-6708/2008/06/030}{\emph{JHEP} {\bf 06}
  (2008) 030}, [\href{http://arxiv.org/abs/0803.3964}{{\tt 0803.3964}}].

\bibitem{Mastrolia:2010nb}
P.~Mastrolia, G.~Ossola, T.~Reiter and F.~Tramontano, \emph{{Scattering
  AMplitudes from Unitarity-based Reduction Algorithm at the Integrand-level}},
  \href{http://dx.doi.org/10.1007/JHEP08(2010)080}{\emph{JHEP} {\bf 08} (2010)
  080}, [\href{http://arxiv.org/abs/1006.0710}{{\tt 1006.0710}}].

\bibitem{Cullen:2011ac}
G.~Cullen, N.~Greiner, G.~Heinrich, G.~Luisoni, P.~Mastrolia, G.~Ossola et~al.,
  \emph{{Automated One-Loop Calculations with GoSam}},
  \href{http://dx.doi.org/10.1140/epjc/s10052-012-1889-1}{\emph{Eur. Phys. J.}
  {\bf C72} (2012) 1889}, [\href{http://arxiv.org/abs/1111.2034}{{\tt
  1111.2034}}].

\bibitem{Bern:1994zx}
Z.~Bern, L.~J. Dixon, D.~C. Dunbar and D.~A. Kosower, \emph{{One loop n point
  gauge theory amplitudes, unitarity and collinear limits}},
  \href{http://dx.doi.org/10.1016/0550-3213(94)90179-1}{\emph{Nucl. Phys.} {\bf
  B425} (1994) 217--260}, [\href{http://arxiv.org/abs/hep-ph/9403226}{{\tt
  hep-ph/9403226}}].

\bibitem{Bern:1994cg}
Z.~Bern, L.~J. Dixon, D.~C. Dunbar and D.~A. Kosower, \emph{{Fusing gauge
  theory tree amplitudes into loop amplitudes}},
  \href{http://dx.doi.org/10.1016/0550-3213(94)00488-Z}{\emph{Nucl. Phys.} {\bf
  B435} (1995) 59--101}, [\href{http://arxiv.org/abs/hep-ph/9409265}{{\tt
  hep-ph/9409265}}].

\bibitem{Bern:1997sc}
Z.~Bern, L.~J. Dixon and D.~A. Kosower, \emph{{One loop amplitudes for e+ e- to
  four partons}},
  \href{http://dx.doi.org/10.1016/S0550-3213(97)00703-7}{\emph{Nucl. Phys.}
  {\bf B513} (1998) 3--86}, [\href{http://arxiv.org/abs/hep-ph/9708239}{{\tt
  hep-ph/9708239}}].

\bibitem{Britto:2004nc}
R.~Britto, F.~Cachazo and B.~Feng, \emph{{Generalized unitarity and one-loop
  amplitudes in N=4 super-Yang-Mills}},
  \href{http://dx.doi.org/10.1016/j.nuclphysb.2005.07.014}{\emph{Nucl. Phys.}
  {\bf B725} (2005) 275--305}, [\href{http://arxiv.org/abs/hep-th/0412103}{{\tt
  hep-th/0412103}}].

\bibitem{Forde:2007mi}
D.~Forde, \emph{{Direct extraction of one-loop integral coefficients}},
  \href{http://dx.doi.org/10.1103/PhysRevD.75.125019}{\emph{Phys. Rev.} {\bf
  D75} (2007) 125019}, [\href{http://arxiv.org/abs/0704.1835}{{\tt
  0704.1835}}].

\bibitem{Ellis:2007br}
R.~K. Ellis, W.~T. Giele and Z.~Kunszt, \emph{{A Numerical Unitarity Formalism
  for Evaluating One-Loop Amplitudes}},
  \href{http://dx.doi.org/10.1088/1126-6708/2008/03/003}{\emph{JHEP} {\bf 03}
  (2008) 003}, [\href{http://arxiv.org/abs/0708.2398}{{\tt 0708.2398}}].

\bibitem{Giele:2008ve}
W.~T. Giele, Z.~Kunszt and K.~Melnikov, \emph{{Full one-loop amplitudes from
  tree amplitudes}},
  \href{http://dx.doi.org/10.1088/1126-6708/2008/04/049}{\emph{JHEP} {\bf 04}
  (2008) 049}, [\href{http://arxiv.org/abs/0801.2237}{{\tt 0801.2237}}].

\bibitem{Berger:2008sj}
C.~F. Berger, Z.~Bern, L.~J. Dixon, F.~Febres~Cordero, D.~Forde, H.~Ita et~al.,
  \emph{{An Automated Implementation of On-Shell Methods for One-Loop
  Amplitudes}}, \href{http://dx.doi.org/10.1103/PhysRevD.78.036003}{\emph{Phys.
  Rev.} {\bf D78} (2008) 036003}, [\href{http://arxiv.org/abs/0803.4180}{{\tt
  0803.4180}}].

\bibitem{Mastrolia:2011pr}
P.~Mastrolia and G.~Ossola, \emph{{On the Integrand-Reduction Method for
  Two-Loop Scattering Amplitudes}},
  \href{http://dx.doi.org/10.1007/JHEP11(2011)014}{\emph{JHEP} {\bf 11} (2011)
  014}, [\href{http://arxiv.org/abs/1107.6041}{{\tt 1107.6041}}].

\bibitem{Kosower:2011ty}
D.~A. Kosower and K.~J. Larsen, \emph{{Maximal Unitarity at Two Loops}},
  \href{http://dx.doi.org/10.1103/PhysRevD.85.045017}{\emph{Phys. Rev.} {\bf
  D85} (2012) 045017}, [\href{http://arxiv.org/abs/1108.1180}{{\tt
  1108.1180}}].

\bibitem{Badger:2012dp}
S.~Badger, H.~Frellesvig and Y.~Zhang, \emph{{Hepta-Cuts of Two-Loop Scattering
  Amplitudes}}, \href{http://dx.doi.org/10.1007/JHEP04(2012)055}{\emph{JHEP}
  {\bf 04} (2012) 055}, [\href{http://arxiv.org/abs/1202.2019}{{\tt
  1202.2019}}].

\bibitem{Zhang:2012ce}
Y.~Zhang, \emph{{Integrand-Level Reduction of Loop Amplitudes by Computational
  Algebraic Geometry Methods}},
  \href{http://dx.doi.org/10.1007/JHEP09(2012)042}{\emph{JHEP} {\bf 09} (2012)
  042}, [\href{http://arxiv.org/abs/1205.5707}{{\tt 1205.5707}}].

\bibitem{Badger:2012dv}
S.~Badger, H.~Frellesvig and Y.~Zhang, \emph{{An Integrand Reconstruction
  Method for Three-Loop Amplitudes}},
  \href{http://dx.doi.org/10.1007/JHEP08(2012)065}{\emph{JHEP} {\bf 08} (2012)
  065}, [\href{http://arxiv.org/abs/1207.2976}{{\tt 1207.2976}}].

\bibitem{Mastrolia:2012an}
P.~Mastrolia, E.~Mirabella, G.~Ossola and T.~Peraro, \emph{{Scattering
  Amplitudes from Multivariate Polynomial Division}},
  \href{http://dx.doi.org/10.1016/j.physletb.2012.09.053}{\emph{Phys. Lett.}
  {\bf B718} (2012) 173--177}, [\href{http://arxiv.org/abs/1205.7087}{{\tt
  1205.7087}}].

\bibitem{Mastrolia:2012wf}
P.~Mastrolia, E.~Mirabella, G.~Ossola and T.~Peraro, \emph{{Integrand-Reduction
  for Two-Loop Scattering Amplitudes through Multivariate Polynomial
  Division}}, \href{http://dx.doi.org/10.1103/PhysRevD.87.085026}{\emph{Phys.
  Rev.} {\bf D87} (2013) 085026}, [\href{http://arxiv.org/abs/1209.4319}{{\tt
  1209.4319}}].

\bibitem{Mastrolia:2013kca}
P.~Mastrolia, E.~Mirabella, G.~Ossola and T.~Peraro, \emph{{Multiloop Integrand
  Reduction for Dimensionally Regulated Amplitudes}},
  \href{http://dx.doi.org/10.1016/j.physletb.2013.10.066}{\emph{Phys. Lett.}
  {\bf B727} (2013) 532--535}, [\href{http://arxiv.org/abs/1307.5832}{{\tt
  1307.5832}}].

\bibitem{CaronHuot:2012ab}
S.~Caron-Huot and K.~J. Larsen, \emph{{Uniqueness of two-loop master
  contours}}, \href{http://dx.doi.org/10.1007/JHEP10(2012)026}{\emph{JHEP} {\bf
  10} (2012) 026}, [\href{http://arxiv.org/abs/1205.0801}{{\tt 1205.0801}}].

\bibitem{Mastrolia:2016dhn}
P.~Mastrolia, T.~Peraro and A.~Primo, \emph{{Adaptive Integrand Decomposition
  in parallel and orthogonal space}},
  \href{http://dx.doi.org/10.1007/JHEP08(2016)164}{\emph{JHEP} {\bf 08} (2016)
  164}, [\href{http://arxiv.org/abs/1605.03157}{{\tt 1605.03157}}].

\bibitem{Peraro:2016wsq}
T.~Peraro, \emph{{Scattering amplitudes over finite fields and multivariate
  functional reconstruction}},
  \href{http://dx.doi.org/10.1007/JHEP12(2016)030}{\emph{JHEP} {\bf 12} (2016)
  030}, [\href{http://arxiv.org/abs/1608.01902}{{\tt 1608.01902}}].

\bibitem{Abreu:2017idw}
S.~Abreu, F.~Febres~Cordero, H.~Ita, M.~Jaquier and B.~Page, \emph{{Subleading
  Poles in the Numerical Unitarity Method at Two Loops}},
  \href{http://dx.doi.org/10.1103/PhysRevD.95.096011}{\emph{Phys. Rev.} {\bf
  D95} (2017) 096011}, [\href{http://arxiv.org/abs/1703.05255}{{\tt
  1703.05255}}].

\bibitem{Badger:2013gxa}
S.~Badger, H.~Frellesvig and Y.~Zhang, \emph{{A Two-Loop Five-Gluon Helicity
  Amplitude in QCD}},
  \href{http://dx.doi.org/10.1007/JHEP12(2013)045}{\emph{JHEP} {\bf 12} (2013)
  045}, [\href{http://arxiv.org/abs/1310.1051}{{\tt 1310.1051}}].

\bibitem{Badger:2015lda}
S.~Badger, G.~Mogull, A.~Ochirov and D.~O'Connell, \emph{{A Complete Two-Loop,
  Five-Gluon Helicity Amplitude in Yang-Mills Theory}},
  \href{http://dx.doi.org/10.1007/JHEP10(2015)064}{\emph{JHEP} {\bf 10} (2015)
  064}, [\href{http://arxiv.org/abs/1507.08797}{{\tt 1507.08797}}].

\bibitem{Dunbar:2016aux}
D.~C. Dunbar and W.~B. Perkins, \emph{{Two-loop five-point all plus helicity
  Yang-Mills amplitude}},
  \href{http://dx.doi.org/10.1103/PhysRevD.93.085029}{\emph{Phys. Rev.} {\bf
  D93} (2016) 085029}, [\href{http://arxiv.org/abs/1603.07514}{{\tt
  1603.07514}}].

\bibitem{Dunbar:2016cxp}
D.~C. Dunbar, G.~R. Jehu and W.~B. Perkins, \emph{{The two-loop n-point
  all-plus helicity amplitude}},
  \href{http://dx.doi.org/10.1103/PhysRevD.93.125006}{\emph{Phys. Rev.} {\bf
  D93} (2016) 125006}, [\href{http://arxiv.org/abs/1604.06631}{{\tt
  1604.06631}}].

\bibitem{Dunbar:2016gjb}
D.~C. Dunbar, G.~R. Jehu and W.~B. Perkins, \emph{{Two-loop six gluon all plus
  helicity amplitude}},
  \href{http://dx.doi.org/10.1103/PhysRevLett.117.061602}{\emph{Phys. Rev.
  Lett.} {\bf 117} (2016) 061602}, [\href{http://arxiv.org/abs/1605.06351}{{\tt
  1605.06351}}].

\bibitem{Badger:2016ozq}
S.~Badger, G.~Mogull and T.~Peraro, \emph{{Local integrands for two-loop
  all-plus Yang-Mills amplitudes}},
  \href{http://dx.doi.org/10.1007/JHEP08(2016)063}{\emph{JHEP} {\bf 08} (2016)
  063}, [\href{http://arxiv.org/abs/1606.02244}{{\tt 1606.02244}}].

\bibitem{Dunbar:2017nfy}
D.~C. Dunbar, J.~H. Godwin, G.~R. Jehu and W.~B. Perkins, \emph{{Analytic
  all-plus gluon amplitudes in QCD}},
  \href{http://arxiv.org/abs/1710.10071}{{\tt 1710.10071}}.

\bibitem{Badger:2017jhb}
S.~Badger, C.~Brønnum-Hansen, H.~B. Hartanto and T.~Peraro, \emph{{First look
  at two-loop five-gluon scattering in QCD}},
  \href{http://dx.doi.org/10.1103/PhysRevLett.120.092001}{\emph{Phys. Rev.
  Lett.} {\bf 120} (2018) 092001}, [\href{http://arxiv.org/abs/1712.02229}{{\tt
  1712.02229}}].

\bibitem{Abreu:2017hqn}
S.~Abreu, F.~Febres~Cordero, H.~Ita, B.~Page and M.~Zeng, \emph{{Planar
  Two-Loop Five-Gluon Amplitudes from Numerical Unitarity}},
  \href{http://dx.doi.org/10.1103/PhysRevD.97.116014}{\emph{Phys. Rev.} {\bf
  D97} (2018) 116014}, [\href{http://arxiv.org/abs/1712.03946}{{\tt
  1712.03946}}].

\bibitem{Badger:2018gip}
S.~Badger, C.~Brønnum-Hansen, T.~Gehrmann, H.~B. Hartanto, J.~Henn, N.~A.
  Lo~Presti et~al., \emph{{Applications of integrand reduction to two-loop
  five-point scattering amplitudes in QCD}},
  \href{http://dx.doi.org/10.22323/1.303.0006}{\emph{PoS} {\bf LL2018} (2018)
  006}, [\href{http://arxiv.org/abs/1807.09709}{{\tt 1807.09709}}].

\bibitem{Abreu:2018jgq}
S.~Abreu, F.~Febres~Cordero, H.~Ita, B.~Page and V.~Sotnikov, \emph{{Planar
  Two-Loop Five-Parton Amplitudes from Numerical Unitarity}},
  \href{http://arxiv.org/abs/1809.09067}{{\tt 1809.09067}}.

\bibitem{Boels:2018nrr}
R.~H. Boels, Q.~Jin and H.~Luo, \emph{{Efficient integrand reduction for
  particles with spin}},  \href{http://arxiv.org/abs/1802.06761}{{\tt
  1802.06761}}.

\bibitem{Chawdhry:2018awn}
H.~A. Chawdhry, M.~A. Lim and A.~Mitov, \emph{{Two-loop five-point massless QCD
  amplitudes within the IBP approach}},
  \href{http://arxiv.org/abs/1805.09182}{{\tt 1805.09182}}.

\bibitem{Tkachov:1981wb}
F.~V. Tkachov, \emph{{A Theorem on Analytical Calculability of Four Loop
  Renormalization Group Functions}},
  \href{http://dx.doi.org/10.1016/0370-2693(81)90288-4}{\emph{Phys. Lett.} {\bf
  100B} (1981) 65--68}.

\bibitem{Chetyrkin:1981qh}
K.~G. Chetyrkin and F.~V. Tkachov, \emph{{Integration by Parts: The Algorithm
  to Calculate beta Functions in 4 Loops}},
  \href{http://dx.doi.org/10.1016/0550-3213(81)90199-1}{\emph{Nucl. Phys.} {\bf
  B192} (1981) 159--204}.

\bibitem{Gluza:2010ws}
J.~Gluza, K.~Kajda and D.~A. Kosower, \emph{{Towards a Basis for Planar
  Two-Loop Integrals}},
  \href{http://dx.doi.org/10.1103/PhysRevD.83.045012}{\emph{Phys. Rev.} {\bf
  D83} (2011) 045012}, [\href{http://arxiv.org/abs/1009.0472}{{\tt
  1009.0472}}].

\bibitem{Ita:2015tya}
H.~Ita, \emph{{Two-loop Integrand Decomposition into Master Integrals and
  Surface Terms}},
  \href{http://dx.doi.org/10.1103/PhysRevD.94.116015}{\emph{Phys. Rev.} {\bf
  D94} (2016) 116015}, [\href{http://arxiv.org/abs/1510.05626}{{\tt
  1510.05626}}].

\bibitem{Larsen:2015ped}
K.~J. Larsen and Y.~Zhang, \emph{{Integration-by-parts reductions from
  unitarity cuts and algebraic geometry}},
  \href{http://dx.doi.org/10.1103/PhysRevD.93.041701}{\emph{Phys. Rev.} {\bf
  D93} (2016) 041701}, [\href{http://arxiv.org/abs/1511.01071}{{\tt
  1511.01071}}].

\bibitem{Georgoudis:2016wff}
A.~Georgoudis, K.~J. Larsen and Y.~Zhang, \emph{{Azurite: An algebraic geometry
  based package for finding bases of loop integrals}},
  \href{http://dx.doi.org/10.1016/j.cpc.2017.08.013}{\emph{Comput. Phys.
  Commun.} {\bf 221} (2017) 203--215},
  [\href{http://arxiv.org/abs/1612.04252}{{\tt 1612.04252}}].

\bibitem{Kosower:2018obg}
D.~A. Kosower, \emph{{Direct Solution of Integration-by-Parts Systems}},
  \href{http://arxiv.org/abs/1804.00131}{{\tt 1804.00131}}.

\bibitem{Boehm:2018fpv}
J.~Böhm, A.~Georgoudis, K.~J. Larsen, H.~Schönemann and Y.~Zhang,
  \emph{{Complete integration-by-parts reductions of the non-planar hexagon-box
  via module intersections}},
  \href{http://dx.doi.org/10.1007/JHEP09(2018)024}{\emph{JHEP} {\bf 09} (2018)
  024}, [\href{http://arxiv.org/abs/1805.01873}{{\tt 1805.01873}}].

\bibitem{Boehm:2017wjc}
J.~Böhm, A.~Georgoudis, K.~J. Larsen, M.~Schulze and Y.~Zhang, \emph{{Complete
  sets of logarithmic vector fields for integration-by-parts identities of
  Feynman integrals}},
  \href{http://dx.doi.org/10.1103/PhysRevD.98.025023}{\emph{Phys. Rev.} {\bf
  D98} (2018) 025023}, [\href{http://arxiv.org/abs/1712.09737}{{\tt
  1712.09737}}].

\bibitem{Laporta:2001dd}
S.~Laporta, \emph{{High precision calculation of multiloop Feynman integrals by
  difference equations}},
  \href{http://dx.doi.org/10.1016/S0217-751X(00)00215-7,
  10.1142/S0217751X00002157}{\emph{Int. J. Mod. Phys.} {\bf A15} (2000)
  5087--5159}, [\href{http://arxiv.org/abs/hep-ph/0102033}{{\tt
  hep-ph/0102033}}].

\bibitem{vonManteuffel:2012np}
A.~von Manteuffel and C.~Studerus, \emph{{Reduze 2 - Distributed Feynman
  Integral Reduction}},  \href{http://arxiv.org/abs/1201.4330}{{\tt
  1201.4330}}.

\bibitem{Smirnov:2014hma}
A.~V. Smirnov, \emph{{FIRE5: a C++ implementation of Feynman Integral
  REduction}}, \href{http://dx.doi.org/10.1016/j.cpc.2014.11.024}{\emph{Comput.
  Phys. Commun.} {\bf 189} (2015) 182--191},
  [\href{http://arxiv.org/abs/1408.2372}{{\tt 1408.2372}}].

\bibitem{Maierhoefer:2017hyi}
P.~Maierhoefer, J.~Usovitsch and P.~Uwer, \emph{{Kira - A Feynman Integral
  Reduction Program}},  \href{http://arxiv.org/abs/1705.05610}{{\tt
  1705.05610}}.

\bibitem{vonManteuffel:2014ixa}
A.~von Manteuffel and R.~M. Schabinger, \emph{{A novel approach to integration
  by parts reduction}},
  \href{http://dx.doi.org/10.1016/j.physletb.2015.03.029}{\emph{Phys. Lett.}
  {\bf B744} (2015) 101--104}, [\href{http://arxiv.org/abs/1406.4513}{{\tt
  1406.4513}}].

\bibitem{Kotikov:1990kg}
A.~V. Kotikov, \emph{{Differential equations method: New technique for massive
  Feynman diagrams calculation}},
  \href{http://dx.doi.org/10.1016/0370-2693(91)90413-K}{\emph{Phys. Lett.} {\bf
  B254} (1991) 158--164}.

\bibitem{Gehrmann:1999as}
T.~Gehrmann and E.~Remiddi, \emph{{Differential equations for two loop four
  point functions}},
  \href{http://dx.doi.org/10.1016/S0550-3213(00)00223-6}{\emph{Nucl. Phys.}
  {\bf B580} (2000) 485--518}, [\href{http://arxiv.org/abs/hep-ph/9912329}{{\tt
  hep-ph/9912329}}].

\bibitem{Henn:2013pwa}
J.~M. Henn, \emph{{Multiloop integrals in dimensional regularization made
  simple}}, \href{http://dx.doi.org/10.1103/PhysRevLett.110.251601}{\emph{Phys.
  Rev. Lett.} {\bf 110} (2013) 251601},
  [\href{http://arxiv.org/abs/1304.1806}{{\tt 1304.1806}}].

\bibitem{Papadopoulos:2014lla}
C.~G. Papadopoulos, \emph{{Simplified differential equations approach for
  Master Integrals}},
  \href{http://dx.doi.org/10.1007/JHEP07(2014)088}{\emph{JHEP} {\bf 07} (2014)
  088}, [\href{http://arxiv.org/abs/1401.6057}{{\tt 1401.6057}}].

\bibitem{vonManteuffel:2014qoa}
A.~von Manteuffel, E.~Panzer and R.~M. Schabinger, \emph{{A quasi-finite basis
  for multi-loop Feynman integrals}},
  \href{http://dx.doi.org/10.1007/JHEP02(2015)120}{\emph{JHEP} {\bf 02} (2015)
  120}, [\href{http://arxiv.org/abs/1411.7392}{{\tt 1411.7392}}].

\bibitem{Gehrmann:2015bfy}
T.~Gehrmann, J.~M. Henn and N.~A. Lo~Presti, \emph{{Analytic form of the
  two-loop planar five-gluon all-plus-helicity amplitude in QCD}},
  \href{http://dx.doi.org/10.1103/PhysRevLett.116.189903,
  10.1103/PhysRevLett.116.062001}{\emph{Phys. Rev. Lett.} {\bf 116} (2016)
  062001}, [\href{http://arxiv.org/abs/1511.05409}{{\tt 1511.05409}}].

\bibitem{Papadopoulos:2015jft}
C.~G. Papadopoulos, D.~Tommasini and C.~Wever, \emph{{The Pentabox Master
  Integrals with the Simplified Differential Equations approach}},
  \href{http://dx.doi.org/10.1007/JHEP04(2016)078}{\emph{JHEP} {\bf 04} (2016)
  078}, [\href{http://arxiv.org/abs/1511.09404}{{\tt 1511.09404}}].

\bibitem{Zeng:2017ipr}
M.~Zeng, \emph{{Differential equations on unitarity cut surfaces}},
  \href{http://dx.doi.org/10.1007/JHEP06(2017)121}{\emph{JHEP} {\bf 06} (2017)
  121}, [\href{http://arxiv.org/abs/1702.02355}{{\tt 1702.02355}}].

\bibitem{Chicherin:2017dob}
D.~Chicherin, J.~Henn and V.~Mitev, \emph{{Bootstrapping pentagon functions}},
  \href{http://dx.doi.org/10.1007/JHEP05(2018)164}{\emph{JHEP} {\bf 05} (2018)
  164}, [\href{http://arxiv.org/abs/1712.09610}{{\tt 1712.09610}}].

\bibitem{Gehrmann:2018yef}
T.~Gehrmann, J.~M. Henn and N.~A. Lo~Presti, \emph{{Pentagon functions for
  massless planar scattering amplitudes}},
  \href{http://dx.doi.org/10.1007/JHEP10(2018)103}{\emph{JHEP} {\bf 10} (2018)
  103}, [\href{http://arxiv.org/abs/1807.09812}{{\tt 1807.09812}}].

\bibitem{Abreu:2018rcw}
S.~Abreu, B.~Page and M.~Zeng, \emph{{Differential equations from unitarity
  cuts: nonplanar hexa-box integrals}},
  \href{http://arxiv.org/abs/1807.11522}{{\tt 1807.11522}}.

\bibitem{Chicherin:2018mue}
D.~Chicherin, T.~Gehrmann, J.~M. Henn, N.~A. Lo~Presti, V.~Mitev and P.~Wasser,
  \emph{{Analytic result for the nonplanar hexa-box integrals}},
  \href{http://arxiv.org/abs/1809.06240}{{\tt 1809.06240}}.

\bibitem{Bern:1993mq}
Z.~Bern, L.~J. Dixon and D.~A. Kosower, \emph{{One loop corrections to five
  gluon amplitudes}},
  \href{http://dx.doi.org/10.1103/PhysRevLett.70.2677}{\emph{Phys. Rev. Lett.}
  {\bf 70} (1993) 2677--2680}, [\href{http://arxiv.org/abs/hep-ph/9302280}{{\tt
  hep-ph/9302280}}].

\bibitem{Caron-Huot:2018dsv}
S.~Caron-Huot, L.~J. Dixon, M.~von Hippel, A.~J. McLeod and G.~Papathanasiou,
  \emph{{The Double Pentaladder Integral to All Orders}},
  \href{http://dx.doi.org/10.1007/JHEP07(2018)170}{\emph{JHEP} {\bf 07} (2018)
  170}, [\href{http://arxiv.org/abs/1806.01361}{{\tt 1806.01361}}].

\bibitem{Dixon:2016nkn}
L.~J. Dixon, J.~Drummond, T.~Harrington, A.~J. McLeod, G.~Papathanasiou and
  M.~Spradlin, \emph{{Heptagons from the Steinmann Cluster Bootstrap}},
  \href{http://dx.doi.org/10.1007/JHEP02(2017)137}{\emph{JHEP} {\bf 02} (2017)
  137}, [\href{http://arxiv.org/abs/1612.08976}{{\tt 1612.08976}}].

\bibitem{Caron-Huot:2016owq}
S.~Caron-Huot, L.~J. Dixon, A.~McLeod and M.~von Hippel, \emph{{Bootstrapping a
  Five-Loop Amplitude Using Steinmann Relations}},
  \href{http://dx.doi.org/10.1103/PhysRevLett.117.241601}{\emph{Phys. Rev.
  Lett.} {\bf 117} (2016) 241601}, [\href{http://arxiv.org/abs/1609.00669}{{\tt
  1609.00669}}].

\bibitem{Dixon:2015iva}
L.~J. Dixon, M.~von Hippel and A.~J. McLeod, \emph{{The four-loop six-gluon
  NMHV ratio function}},
  \href{http://dx.doi.org/10.1007/JHEP01(2016)053}{\emph{JHEP} {\bf 01} (2016)
  053}, [\href{http://arxiv.org/abs/1509.08127}{{\tt 1509.08127}}].

\bibitem{Almelid:2017qju}
{Almelid, \O{}yvind and Duhr, Claude and Gardi, Einan and McLeod, Andrew and
  White, Chris D.}, \emph{{Bootstrapping the QCD soft anomalous dimension}},
  \href{http://dx.doi.org/10.1007/JHEP09(2017)073}{\emph{JHEP} {\bf 09} (2017)
  073}, [\href{http://arxiv.org/abs/1706.10162}{{\tt 1706.10162}}].

\bibitem{Catani:1998bh}
S.~Catani, \emph{{The Singular behavior of QCD amplitudes at two loop order}},
  \href{http://dx.doi.org/10.1016/S0370-2693(98)00332-3}{\emph{Phys. Lett.}
  {\bf B427} (1998) 161--171}, [\href{http://arxiv.org/abs/hep-ph/9802439}{{\tt
  hep-ph/9802439}}].

\bibitem{Becher:2009qa}
T.~Becher and M.~Neubert, \emph{{On the Structure of Infrared Singularities of
  Gauge-Theory Amplitudes}},
  \href{http://dx.doi.org/10.1088/1126-6708/2009/06/081,
  10.1007/JHEP11(2013)024}{\emph{JHEP} {\bf 06} (2009) 081},
  [\href{http://arxiv.org/abs/0903.1126}{{\tt 0903.1126}}].

\bibitem{Becher:2009cu}
T.~Becher and M.~Neubert, \emph{{Infrared singularities of scattering
  amplitudes in perturbative QCD}},
  \href{http://dx.doi.org/10.1103/PhysRevLett.102.162001,
  10.1103/PhysRevLett.111.199905}{\emph{Phys. Rev. Lett.} {\bf 102} (2009)
  162001}, [\href{http://arxiv.org/abs/0901.0722}{{\tt 0901.0722}}].

\bibitem{Gardi:2009qi}
E.~Gardi and L.~Magnea, \emph{{Factorization constraints for soft anomalous
  dimensions in QCD scattering amplitudes}},
  \href{http://dx.doi.org/10.1088/1126-6708/2009/03/079}{\emph{JHEP} {\bf 03}
  (2009) 079}, [\href{http://arxiv.org/abs/0901.1091}{{\tt 0901.1091}}].

\bibitem{Hodges:2009hk}
A.~Hodges, \emph{{Eliminating spurious poles from gauge-theoretic amplitudes}},
  \href{http://dx.doi.org/10.1007/JHEP05(2013)135}{\emph{JHEP} {\bf 05} (2013)
  135}, [\href{http://arxiv.org/abs/0905.1473}{{\tt 0905.1473}}].

\bibitem{ArkaniHamed:2010kv}
N.~Arkani-Hamed, J.~L. Bourjaily, F.~Cachazo, S.~Caron-Huot and J.~Trnka,
  \emph{{The All-Loop Integrand For Scattering Amplitudes in Planar N=4 SYM}},
  \href{http://dx.doi.org/10.1007/JHEP01(2011)041}{\emph{JHEP} {\bf 01} (2011)
  041}, [\href{http://arxiv.org/abs/1008.2958}{{\tt 1008.2958}}].

\bibitem{ArkaniHamed:2010gh}
N.~Arkani-Hamed, J.~L. Bourjaily, F.~Cachazo and J.~Trnka, \emph{{Local
  Integrals for Planar Scattering Amplitudes}},
  \href{http://dx.doi.org/10.1007/JHEP06(2012)125}{\emph{JHEP} {\bf 06} (2012)
  125}, [\href{http://arxiv.org/abs/1012.6032}{{\tt 1012.6032}}].

\bibitem{Bourjaily:2017wjl}
J.~L. Bourjaily, E.~Herrmann and J.~Trnka, \emph{{Prescriptive Unitarity}},
  \href{http://dx.doi.org/10.1007/JHEP06(2017)059}{\emph{JHEP} {\bf 06} (2017)
  059}, [\href{http://arxiv.org/abs/1704.05460}{{\tt 1704.05460}}].

\bibitem{Berends:1987me}
F.~A. Berends and W.~T. Giele, \emph{{Recursive Calculations for Processes with
  n Gluons}}, \href{http://dx.doi.org/10.1016/0550-3213(88)90442-7}{\emph{Nucl.
  Phys.} {\bf B306} (1988) 759--808}.

\bibitem{Cheung:2009dc}
C.~Cheung and D.~O'Connell, \emph{{Amplitudes and Spinor-Helicity in Six
  Dimensions}},
  \href{http://dx.doi.org/10.1088/1126-6708/2009/07/075}{\emph{JHEP} {\bf 07}
  (2009) 075}, [\href{http://arxiv.org/abs/0902.0981}{{\tt 0902.0981}}].

\bibitem{Lee:2012cn}
R.~N. Lee, \emph{{Presenting LiteRed: a tool for the Loop InTEgrals
  REDuction}},  \href{http://arxiv.org/abs/1212.2685}{{\tt 1212.2685}}.

\bibitem{Bern:1994fz}
Z.~Bern, L.~J. Dixon and D.~A. Kosower, \emph{{One loop corrections to two
  quark three gluon amplitudes}},
  \href{http://dx.doi.org/10.1016/0550-3213(94)00542-M}{\emph{Nucl. Phys.} {\bf
  B437} (1995) 259--304}, [\href{http://arxiv.org/abs/hep-ph/9409393}{{\tt
  hep-ph/9409393}}].

\end{thebibliography}\endgroup

\end{document}